\documentclass[prd,aps,twocolumn,nofootinbib,preprintnumbers]{revtex4}
\usepackage{amsmath,amssymb,epsf}

\begin{document}

\title { \Large  \bf  BRANECODE\\A Program for Simulations
of\\Braneworld Dynamics}
\author{Johannes Martin$^{1}$, Gary N. Felder $^2$, Andrei V. Frolov $^3$,
Lev Kofman $^1$, and Marco Peloso $^4$}
\affiliation{${}^1$CITA, University of Toronto, 60 St George St,
Toronto, ON M5S 3H8, Canada}
\affiliation{${}^2$Department of Physics, Clark Science Center, Smith
College Northampton, MA 01063, USA}
\affiliation{${}^3$KIPAC/ITP, Stanford University, Stanford, CA 94305-4060, USA}
\affiliation{${}^4$School of Physics and Astronomy, University of
Minnesota, Minneapolis, MN 55455, USA.}
\date{\today}

\begin{abstract}
  We describe an algorithm and a C++ implementation that we have
  written and made available for calculating the fully nonlinear
  evolution of $5$D braneworld models with scalar fields. Bulk fields
  allow for the stabilization of the extra dimension.  However, they
  complicate the dynamics of the system, so that analytic calculations
  (performed within an effective $4$D theory) are usually only
  reliable for static bulk configurations or when the evolution of the
  extra dimension is negligible. In the general case, the nonlinear
  $5$D dynamics can be studied numerically, and the algorithm and code
  we describe are the first ones of that type designed for this task.
  The program and its full documentation are available on the Web at
  \texttt{http://www.cita.utoronto.ca/{\textasciitilde}jmartin/BRANECODE/}~\footnote{We also maintain a mirror of the BRANECODE website at 
    \texttt{http://www.cita.utoronto.ca/{\textasciitilde}kofman/BRANECODE/}}.
  In this paper we provide a brief overview of what the program does
  and how to use it.
\end{abstract}

\preprint{CITA-2004-14}
\preprint{SU-ITP-04-14}
\preprint{hep-ph/0404141}

\maketitle

\section{Introduction}

Many extensions of the Standard Model have in common the presence of
extra dimensions. This has to be contrasted with the fact that our
world looks four dimensional, so one has to explain why the presence
of the extra space has not yet been detected. The traditional answer
has been that the extra space is compact and very small, so that the
fields associated with its excitations are too heavy to be observable
in accelerators or cosmology. More recently, it has been realized that
ordinary matter and gauge interactions may be confined on lower
dimensional submanifolds, known as branes. In this case, they could be
four dimensional objects, even if the geometry of the theory is higher
dimensional. The situation is different for gravity, which propagates
in the whole bulk space. Several questions naturally arise, such as
why a compact space would remain small while the three non-compact
dimensions are undergoing cosmological expansion, or why the expansion
of the universe we see is described by $3+1$ dimensional general
relativity so well. The presence of extra dimensions may 
cause  deviations from the standard FRW cosmology that is
 supported by observations.

In most cases, these two questions turn out to be intimately related.
Only if the extra space is static can the evolution of the non-compact
coordinates behave as in the standard four dimensional case. Hence, the
dynamics of the hidden dimensions becomes a crucial ingredient in
understanding the evolution of the ones we observe. In some particular
cases, static bulk configurations can be achieved under the combined
action of the bulk/brane gravity. In most realistic examples that
could account for our observed four  dimensional cosmology the stability
is due to the presence of additional fields that acquire nontrivial
configurations in the bulk. While the stabilization has to be
effective at relatively ``late'' times, the first stages of our
universe (before primordial nucleosynthesis, for instance) are much
less constrained.  The evolution of the bulk may have been significant
at this phase, and this offers many new possibilities for
phenomenology. This is particularly true with the addition of the
fields responsible for the ``late'' time stabilization, since they
constitute new dynamical degrees of freedom for the system.

While the above considerations are valid for all models with extra
dimensions, significant computations have been performed in the
framework of brane models. These models can be thought of as
simplified, phenomenological (bottom-up) versions of branes in string
theory. The string dynamics is ignored and the primary focus is on the
classical dynamics from the point of view of General Relativity.
Branes act as a source for the Einstein equations of the system, with
their tension and possibly with the energy density of fields confined
on them.  Additional sources are a bulk cosmological constant, or the
energy density of possible bulk fields. This set-up is sufficiently
rich to describe very interesting situations. For example, inflation
in braneworlds can acquire a nice geometrical interpretation, with the
inflaton associated with the distance between different branes, the
Hubble parameter scale associated with the induced curvature on the
branes, and with reheating through radion oscillations.

Despite this great simplification, the whole dynamics is still very
complicated, in particular when a bulk scalar field is present. In
this case, analytical computations are typically performed within an
effective $4$ dimensional theory, obtained after the extra dimension
is integrated out, or perturbatively, using linearized analysis around
simple backgrounds. While these studies  are very useful when the
extra space is static or quasi-static, they are not sufficient  to
describe the system when the evolution of the bulk is important. For
example, systems that are stable at low energy (low curvature of the
branes) can become unstable when the energy/curvature is increased.
The stability/instability can be studied analytically. However, one
cannot determine analytically where the system will evolve towards
when the initial configuration turns out to be unstable.

For example, we numerically examined the dynamics of brane collisions
and found that, as the branes approach each other, the spacetime of
the bulk asymptotically  approaches the Kasner-type solution. 

Motivated by these limitations with the analytical treatment, we
undertook a numerical study of these models. We developed a numerical
algorithm, implemented in a C++ code, specifically designed for
codimension one braneworlds, with a scalar field included in order to
provide stabilization of the bulk at low energies. With the assumption
of homogeneity and isotropy along the brane spatial coordinates ${\bf
x}\,$ (corresponding to the standard assumption of homogeneity and
isotropy of the non-compact coordinates), the problem is reduced to an
effectively $2$ dimensional one. The independent variables are the
bulk time $t$ and the bulk dimension $y$.  The program integrates
numerically the full set of Einstein equations in the bulk, together
with the Israel junction conditions at two orbifold branes. We
discretize the two dimensional spacetime (time and the bulk
coordinate) and solve the bulk differential equations by
finite-differencing them using the so-called {\it leapfrog}
scheme. This algorithm is sufficient for our problem, and it provides
a good compromise between accuracy and computational time (see Section
\ref{overview} for more details). The two branes act as (one
dimensional) boundaries of this space, and the junction conditions
provide the boundary conditions for the system at each time-step. The
solution of a boundary value problem is required to provide generic
initial conditions that fulfill the constraint equations and the
boundary conditions at the beginning of the evolution. In the static
configurations we have mentioned above, this boundary value problem is
significantly simpler than in the general case. The setup of initial
conditions is explained in more detail in Section \ref{sec:init}.

The first results obtained with the code have been presented
in~\cite{Martin:2003yh}.  We are now making the code public on the
World Wide Web under the name BRANECODE. Its website is at
\texttt{http://www.cita.utoronto.ca/{\textasciitilde}kofman/BRANECODE/}.
The website for the program has documentation, including derivations
of all the equations used in the program. Here we present a short
summary of what the program does and what it can be used for. For more
details see the website. Section \ref{overview} of this paper gives an
overview of what the program is and how it works.  Sections
\ref{equations} and \ref{sec:init} describe the evolution equations
and the setting of initial conditions respectively. Section
\ref{output} describes some of the output generated by the program.
The references section is limited to papers from our group related to
the BRANECODE design and its first results~\cite{Martin:2003yh,
  Frolov:2003yi, Kofman:2004tk, Contaldi:2004hr, Felder:2001da}. See
these papers for a more complete set of references.

\section{Overview and User Adjustable Files}\label{overview}

In this Section we give an overview of the program and how to adjust
it for a particular simulation. More details can
be found in the documentation. 

To work with the BRANECODE the user must specify a model, consisting
of bulk and brane potentials for the scalar field, plus initial
conditions for the field and the geometry. This information is encoded
in a {\it model file}, which is a header file read in by the program.
The file should be called {{\ttfamily model\_numeric.h}} or
{{\ttfamily model\_analytic.h}} depending on whether the initial
conditions are specified numerically or analytically. The model files
contain the potentials and their first and second derivatives that are
needed for the evolution of the bulk equations \eqref{eq:eom} and
boundary conditions \eqref{eq:bc}. For example, the BRANECODE
distribution includes both a numeric and an analytic model file (with
different initial conditions). The file {{\ttfamily
    model\_analytic.h}} contains examples for branes in AdS and
AdS-Schwarzschild geometries, whereas the file {{\ttfamily
    model\_numeric.h}} we designed for the class of models with bulk
scalar fields determined by the bulk and brane potentials
\begin{eqnarray}
\label{eq:pot}
V(\phi) &=& \tfrac{1}{2} \, m^2 \phi^2 + \Lambda \,\,, \nonumber\\
{U}_i (\phi)&=& \tfrac{1}{2}M_i \left( \phi_i - \sigma_i \right)^2 + \lambda_i \,\,.
\end{eqnarray}
A bulk cosmological constant and brane tensions is included as
constant terms in these potentials. Most importantly, the program is
designed to work for arbitrary potentials, different from
(\ref{eq:pot}). Other potentials $V(\phi)$, $U_i(\phi)$ can be
implemented by modifying the corresponding lines in the model file.
Aside from this file, the only other file that the user needs to
modify is {{\ttfamily parameters.h}}, which contains all the
parameters for a given run of the program.  These include the number
of grid points, the running time, and a number of other general
variables specific to each run. There is also a parameter in this file
that tells the program which type of model file to look for.

Given a specific model and set of parameters, the BRANECODE
solves the system of equations of motions for the metric functions and
the scalar field \eqref{eq:eom} along with the boundary
conditions provided by the presence of the branes \eqref{eq:bc}. The
required functions are contained in the file {{\ttfamily equations.cpp}}.

\begin{figure}[h]
\leavevmode\epsfxsize=\columnwidth \epsfbox{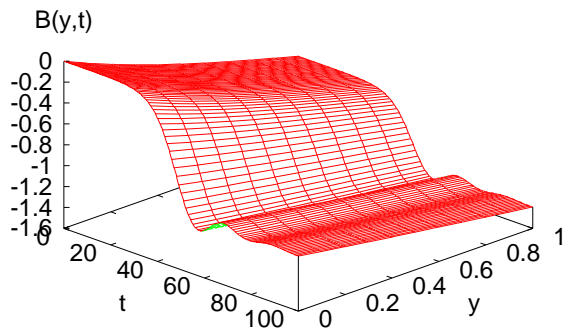}
\caption{An example of the graphical output generated by the
BRANECODE. The plot shows the evolution of a metric component $B$
(describing the interbrane separation) as a function of space and
time. Physically, this evolution shows a transition from an unstable
static warped geometry solution towards a stable static solution.
During this non-linear reconfiguration the interbrane distance and the
Hubble scale of the de Sitter geometry decreases.}
\label{fig:plots}
\end{figure}

The BRANECODE has built-in routines for outputting and plotting the
metric fields, the scalar field, and derived quantities. These outputs
are stored in ASCII files that can be read in and plotted by any
standard plotting software. There are also options (set in {{\ttfamily
parameters.h}}) to have the program call {\ttfamily GNUPLOT} to
generate and display postscript plots of the data at runtime. (These
options should only be chosen if the program is running on a computer
with both {\ttfamily GNUPLOT} and {\ttfamily GHOSTVIEW}.) An example
of this graphical output is shown in Fig.~\ref{fig:plots}.

Once all parameters have been set and you have modified or created a
model file according to your wishes you simply compile and run the
BRANECODE.  The code is designed to be platform independent and should
work with any C++ compiler. The makefile that comes with the
distribution has entries and flags for the GNU \texttt{gcc} compiler
and the INTEL \texttt{icc} compiler.  You can select one of them or
edit the makefile to invoke your favorite compiler.

\section{Algorithm}\label{equations}

The evolution equations solved by the BRANECODE are the set of
Einstein/scalar field equations on an effectively $2$ dimensional
spacetime obtained after imposing homogeneity and isotropy on the
non-compact spatial coordinates ${\bf x}\,$. In~\cite{Martin:2003yh}
we showed that under these conditions it is always possible to choose
coordinates that bring the metric to the form
\begin{equation}
d s^2 ={\rm e}^{2\,B \left( t, y \right)} \left(- d t^2+ d y^2 \right)
 + {\rm e}^{2\,A \left( t, y \right)} \, d{\bf x}^2\,\,,
\label{line}
\end{equation}
with the two branes fixed at $y=0,\,1\,$, respectively. This choice is
motivated by the fact that in this coordinate system the lattice size
is time independent due to the fixed position of the branes and the
equations simplify significantly.  In this gauge, we have the
following dynamical equations
\begin{eqnarray}
\ddot{A\,\,}\! - A'' + 3 \dot{A}^2 - 3 A'^2 &=& \frac{2}{3}\, {\rm e}^{2 B}
V \,\,, \nonumber\\
\ddot{B} -B'' - 3 \dot{A}^2 + 3 A'^2 + \frac{1}{2}\dot{\phi}^2 -
\frac{1}{2}{\phi'}^2 &=& - \frac{1}{3} {\rm e}^{2 B} V \,\,, \\
\ddot{\phi} - \phi'' + 3 \dot{A} \dot{\phi} - 3 A' \phi' &=& - {\rm
e}^{2 B} V_{,\phi} \,\,,\nonumber
\label{eq:eom}
\end{eqnarray}
supplemented by the constraint equations
\begin{eqnarray}
\label{eq:constraint}
- A' \dot{A} + B' \dot{A} + A' \dot{B} - \dot{A}' &=&
\frac{1}{3} \, \dot{\phi} \, \phi' \,\,, \\
2 A'^2 - A' B' + A'' - \dot{A}^2 - \dot{A} \dot{B} &=&
- \frac{1}{6}{\dot{\phi}^2} -
\frac{1}{6}{\phi'^2} - \frac{1}{3} \, {\rm e}^{2 B} V \,\,. \nonumber
\end{eqnarray}
Dots and primes denote derivatives with respect to the time $t$ and 
the coordinate $y\,$ along the bulk, respectively. In addition, the
program imposes the following junction (Israel) conditions at the
positions of the branes
\begin{equation}
\label{eq:bc}
A' = - \frac{1}{6} \, {U} \, {\rm e}^B \;,\;\;
B' = - \frac{1}{6} \, {U} \, {\rm e}^B \;,\;\;
\phi' = \frac{1}{2} {\rm e}^B \, U_{,\phi} \,\,.
\end{equation}
These junction conditions are equivalent to extending the space beyond
the two branes, and imposing ${\mathbb Z}_2$ symmetry across the
branes.

In the coordinate system~(\ref{line}), the characteristic propagation
speed of the dynamical equations is always $1$. This is advantageous
from a numerical viewpoint as the size of the time step can be
optimized uniformly by setting it equal to spatial grid separation.
We describe below our implementation of the {\it leapfrog}
discretization scheme, which is stable, second order accurate, and
non-dissipative.

The program discretizes the $2$ dimensional $\left\{ t,y \right\}$
space and computes the value of the three functions $B,\,A,\,\phi$ at
each grid point. For any fixed time, the grid is made of $N+1$ points
equally spaced along the bulk, with $0$ and $N$ corresponding to the
locations of the two branes. The value of $N$ is set in {{\ttfamily
    parameters.h}}. The same grid spacing is taken in the $y$ and $t$
directions. The initial conditions are in the form $\left( A(y), B(y),
  \phi(y), \dot A(y), \dot B (y),\dot\phi(y) \right)\,$, and they can
be chosen arbitrarily subject to the fact that they
satisfy~(\ref{eq:constraint}) and (\ref{eq:bc}) at the initial time.
Of particular interest are configurations with an initially static
bulk, since the program can be used to verify their stability and to
study the dynamics when they are unstable as described in section
\ref{sec:init}.

The discretized initial data $\left(\rule{0em}{1em} A_i \right .$,
$\dot{A}_i$, $B_i$, $\dot{B}_i$, $\phi_i$ , $\left .\dot{\phi}_i
  \rule{0em}{1em}\right) \;,\;\; i = 0,\dots, N$ has to be converted
to a form suitable for the {\it leapfrog} algorithm.  Instead of having the
fields and their velocities at the time $t=t_0$, the algorithm needs
the spatial profiles of the fields $(A_i, B_i, \phi_i)$ at two
subsequent moments in time $t=t_0$ and $t=t_0+\epsilon$. In some
special cases the profiles at the two initial times can be calculated
analytically, but in general the program takes a Runge-Kutta step
using the initial derivatives to calculate the field values at
$t_0+\epsilon$.

\begin{figure}[h]
\leavevmode\epsfxsize=0.95\columnwidth \epsfbox{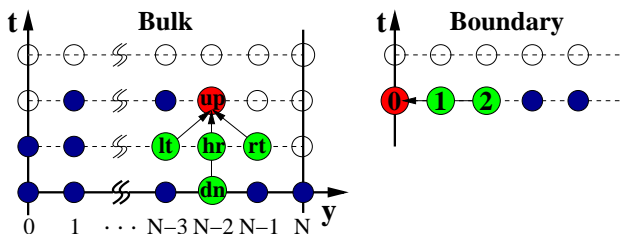}
\caption{Numerical evolution scheme. See text for details.}
\label{fig:num_evolution}
\end{figure}

The field values at all subsequent time steps are calculated as
follows. (Note that we will use $f$ here to denote a generic field,
i.e. for equations that apply to all three fields $A$, $B$, and
$\phi$.) In the gauge chosen, the bulk equations only contain
derivatives in the form $\ddot{f} - f''\,$, and $\dot{f} \, \dot{g} -
f' \, g' \,$. Recall that our grid spacing $\epsilon=1/N$ is equal to
our time step. Thus, to second order accuracy, we can write the
derivatives at a given point in spacetime in terms of the values of
its neighbors as
\begin{eqnarray}
\ddot{f}  - f^{\prime\prime} &=&  \frac{1}{\epsilon^2} \left(  f_{\rm up} +
f_{\rm dn} - f_{\rm lt} - f_{\rm rt}  \right) + {\rm O} \left( \epsilon^2 \right)
\,\, , \nonumber\\
{\dot f} \, {\dot g} - {f^\prime} {g^\prime} &=& \frac{1}{4 \,
\epsilon^2} \left[ \left( f_{\rm up} - f_{\rm dn} \right) \, \left( g_{\rm up} -
g_{\rm dn} \right) + \right. \nonumber\\
&& \left. - \left( f_{\rm rt} - f_{\rm lt} \right) \, \left( g_{\rm rt} -
g_{\rm lt} \right) \right] + {\rm O} \left( \epsilon^2 \right) \,\,,
\label{discrete}
\end{eqnarray}
where the indices label relative grid positions as defined in
Fig.~\ref{fig:num_evolution}. In this way, the three differential
equations~(\ref{eq:eom}) become three algebraic equations for the
three unknown quantities $\left( B_{\rm up} \,, A_{\rm up} \,,
  \phi_{\rm up} \right) \,$. Notice that this can be done
independently site by site in the bulk.

Once the bulk field values at the new time have been determined, we
can apply the junction conditions~(\ref{eq:bc}) to advance the
boundary field values. We discuss only the computation at the first
brane ($i=0\,$). The treatment for the second brane is analogous. Only
first derivatives in $y$ enter in~(\ref{eq:bc}). To preserve second
order accuracy, we use an ``asymmetric'' discretization of the
derivatives
\begin{eqnarray}
f^\prime_0 &=& \frac{1}{2 \, \epsilon} \left( - \, 3 \, f_0 + 4 f_1 -
f_2 \right) + {\rm O} \left( \epsilon^2 \right) \,\,,
\label{eq:bc_difference}
\end{eqnarray}
where subscripts indicate the grid position of each field value (see
Fig.~\ref{fig:num_evolution}). The junction conditions~(\ref{eq:bc})
thus become a set of three algebraic equations in terms of three
unknowns. We can eliminate $A_0$ and $B_0$ in favor of $\phi_0\,$, and
write an equation in terms of the only unknown quantity $\phi_0\,$,
\begin{equation}
4 \, \phi_1 - \phi_2 - 3 \, \phi_0 - \epsilon \, e^{B_0(\phi_0)} \,
{U}_0^\prime(\phi_0) = 0 \,\,.
\label{eq:bc_diff_c}
\end{equation}
For specific brane potentials, these equations can be solved
analytically.  However, as the code is designed for arbitrary
bulk/brane potentials, eq.~(\ref{eq:bc_diff_c}) is solved numerically
using the iterative Newton's method. Once $\phi_0$ is determined, the
remaining unknowns $B_0$ and $A_0$ are trivially computed through the
remaining junction conditions.

\section{Initial conditions}
\label{sec:init}

In general, the specification of initial conditions, i.e.\ the
determination of the initial spatial profiles for the fields and their
velocities $A(y)$, $B(y)$, $\phi(y)$, $\dot A(y)$, $\dot B(y)$, and
$\dot\phi(y)$ is a non-trivial task. These function cannot be chosen
independently, but rather are subject to the constraint equations
\eqref{eq:constraint} and boundary conditions \eqref{eq:bc}.  However,
the physical instability of static de Sitter configurations
\cite{Frolov:2003yi, Martin:2003yh} provides the possibility to
generate interesting braneworld dynamics with static solutions as
initial conditions that we can recommend. All the static solutions for
a given model can be exhaustively classified by the phasespace
analysis of the dynamics of the gravity/scalar system performed in
\cite{Felder:2001da}.

One simple way to generate static initial conditions is to consider a
configuration of the form $B = B \left( y \right) \,,\; \phi = \phi
\left( y \right) \,,\; A = B + H \, t \,$ (de Sitter branes in a
static bulk). The first of the equations (\ref{eq:constraint}) is then
trivially satisfied. For a sufficiently simple model, the remaining
constraint equation can be solved analytically.  Otherwise we provide
a {\ttfamily MATHEMATICA} notebook and {\ttfamily MAPLE} worksheet to
solve them numerically for a given bulk potential and generate
appropriately formatted initial data.  For simplicity, the code
determines two of the three parameters of the brane potentials
\eqref{eq:pot} in such a way that the boundary conditions
\eqref{eq:bc} are fulfilled initially.  The third parameter, either
the tension on the brane $\lambda_i$ or the minimum of the brane
potential $\sigma_i$, is set by the user in the file {\ttfamily
  model\_numeric.h}. If the user instead wants to set up initial
conditions for a given choice of brane potentials, he can make use of
a shooting method operating in the phasespace of the static solutions.
Namely, he can start with initial conditions chosen such that they
fulfill the boundary conditions on one of the branes, compute the
corresponding bulk configuration, and keep varying the initial
conditions (e.g.\ with a numerical scan) until the solution also
satisfies the junction conditions at the second boundary. For any
given set of bulk and brane potentials, there can be none, one, or
more than one solution to the boundary value problem. The latter case
opens up the possibility for interesting dynamics of transitions as
investigated in \cite{Martin:2003yh}.

\section{Output}
\label{output}
The main outputs of the program are the values of the three functions
$B ,\, A$, and $\phi$ at different bulk sites and time steps. Two
parameters inside {{\ttfamily parameters.h}} control how many points
(both in the $y$ and $t$ directions) are to be saved. From these
quantities, one can construct some outputs of immediate physical
relevance. One is the physical interbrane distance. The branes are at
a fixed coordinate distance in our gauge and their physical separation
is encoded in the metric coefficient $B\,$.
\begin{equation}
D \left( t \right) \equiv \int_0^1 dy \, \sqrt{g_{55}} = \int_0^1 dy
\, {\rm e }^{B \left( t ,\, y \right)}\ .
\label{d}
\end{equation}
Another interesting quantity is the Hubble parameter as computed by
observers on each of the two branes
\begin{equation}
H_i \equiv \frac{1}{a} \, \frac{d\,a}{d\,\tau} \Big\vert_i = {\rm
e}^{-B_i} \, \dot{A}_i \,\,,
\label{hubble}
\end{equation}
where $\tau$ is the physical time measured by the observers, defined
by $D \tau \equiv {\rm e }^B \, d t \,$. Quite interestingly, the
gauge choice used in the algorithm (see the previous Section) does not
exhaust the gauge freedom of the problem. Residual gauge
transformations have been described in~\cite{Martin:2003yh}, and one
can show that they do not change the values of $D$ and $H_i\,$, which
therefore have physical meaning.  Besides these physical quantities the
BRANECODE also computes the Ricci scalar and the square of the Weyl
tensor in the bulk. Other quantities of interest can easily be
obtained from the ``raw'' values of $B,A,\phi$.

\section{Conclusions}

The main motivation for our work was to extend the knowledge of
braneworld dynamics beyond the few situations where it was known
analytically. Apart from these situations (characterized by a static
or slowly evolving bulk), approximate methods based on effective $4$
dimensional computations are unreliable. In this short note we have
presented an algorithm, together with its C++ implementation, designed
for numerical computations in this framework. First results obtained
from this code were presented in~\cite{Martin:2003yh,Kofman:2004tk}.
We could show that some bulk configurations which are stable at low
energy (low value of the expansion rate $H$ of the two branes) become
unstable as $H$ increases, in agreement with the analytical
calculations of~\cite{Frolov:2003yi}.  This can be interpreted as a
part of a more general phenomenon of gravitational instability of
compactification to four dimensional de Sitter
geometry~\cite{Contaldi:2004hr}.  The numerical integration allowed us
to follow the evolution of the system starting from the unstable
configuration. For certain bulk/brane potentials the system may evolve
towards another static, but stable configuration, characterized by a
lower value of $H$. The transition is typically a process of quick
bulk reconfiguration.  In many cases, however, the second
configuration does not exist or cannot be reached. The two branes then
either move apart to infinite distance, or they collide.  Brane
collisions is a very interesting subject by itself. The study of the
numerical evolution led us to the conclusion that the asymptotic
geometry is given by the universal Kasner-type that is typical of
homogeneous but anisotropic strong gravity regimes. We did not
investigated the case of branes departing from each other when one of
them in general approaches a naked singularity in the bulk
configuration.  We also did not studied the possibility of the
formation of an apparent horizon between the branes.

Our algorithm is focused on the simplest possible set-up which allows
for brane stabilization based on a generalized Goldberger-Wise
mechanism. Only one bulk scalar field has been considered, although
with arbitrary potentials in the bulk and at the two branes. The
inclusion of more scalar fields would be straightforward. In
particular, one could consider other fields, which are confined to the
branes, and which are coupled to the bulk fields e.g. through the
brane potentials. (The interplay between bulk/brane fields may lead to
novel interesting features not considered in~\cite{Martin:2003yh}).
Another easy generalization would be the inclusion of perfect fluids
on the branes (for example, describing standard matter and radiation).
Less trivial but more interesting extensions could be the inclusion of
other types of fields (form fields in the bulk, for instance), or
evolution with more dimensions included. For example, relaxing the
hypothesis of homogeneity and isotropy along the ordinary dimensions
would allow the study of inhomogeneous perturbations of the
system.

\end{document}